\documentclass[aps,prd,eqsecnum,showpacs]{revtex4}
\usepackage[dvips]{color,graphicx}
\usepackage{amsfonts}
\usepackage{amssymb}

\begin{document}

\title{
Reply to Flambaum Commentary on PRL {\bf 99}, 061301 (2007) Black Hole Constraints on Varying Fundamental Constants}

\author{$^{1}$Jane H. MacGibbon\footnote{Electronic address:  jmacgibb@unf.edu} }

\affiliation{
$^{1}$Department of Chemistry and Physics, University of North Florida,
Jacksonville, Florida 32224, USA
}


\begin{center}Physical Review Letters {\bf 102}, 069002 (February 11 2009)
\end{center}

\pacs{04.70.Dy, 06.20.Jr, 97.60.Lf, 98.80.Es
\hfill UNF-Astro-2-11-09} 

\maketitle

In \cite{JHM1}, it was shown by explicit calculation using only Standard Model physics and experimentally-confirmed QED behaviour that the value of  time-variation in the electron charge $e$ which corresponds to the cosmologically-measured variation in the fine structure constant $\alpha = e^2$ claimed by Webb et al.~\cite{W1} does not violate the Generalized Second Law of Thermodynamics \cite{B1} applied to the theoretically-allowed range of black holes in the present Universe. Therefore, the possibility exists that the measurement may be explained by a variation in the `effective' (i.e. measurable) charge of the electron arising in QED and Standard Model physics. Accelerator experiments have established that the `effective' electron charge varies with the interaction energy scale and many time-dependent energy scales are known in the Universe, most notably the isoentropically-cooling cosmic microwave background temperature. Furthermore, it was explicitly shown in \cite{JHM1} that the claimed measured value {\it matches} the maximal variation allowed by the Generalized Second Law of Thermodynamics with Standard Model physics. Thus, if the Webb et al.~measurement is correct, the calculation of \cite{JHM1} could be used to mathematically or physically elucidate the variation mechanism and principles governing entropy in the Universe. 

Alternatively, the variation may arise from physics beyond the Standard Model, as proposed by Flambaum \cite{F1} and others. Any extended model, however, must match Standard Model physics on the experimentally-confirmed scales. The strongest constraint in \cite{JHM1} comes from highly-charged black holes whose temperature $T_{bh}$ equals the cosmic microwave background temperature, mass $M_{bh}$ is $M_{CMB}\approx 4.5 \times 10^{25}$ g and size is much greater than atomic lengths. Thus new physics could only modify the calculation in \cite{JHM1} if it introduces terms significant on large scales. The argument in \cite{F1} hinges on being able to apply Eq (3) of \cite{F1} - that the ratio $\mu$ of $M_{bh}$ to the Planck mass $M_{pl}$ obeys $\mu =  \left( (S_{bh}/\pi) + Z^2 \alpha \right)/2\sqrt{S_{bh}/\pi}$ where the black hole entropy $S_{bh}$ is {\it constant} with respect to time $t$ - to {\it all} $M_{pl}\leq M_{bh} \leq M_{max}$ where $M_{max}$ is at least $10^{10}M_\odot$ from astrophysical observations, and charges $0\leq Z < Z_{max}$ where $Z_{max} = \sqrt{M_{bh}/\alpha}$ is the Reissner-Nordstr\"{o}m maximal charge. That is, \cite{F1} assumes that time variation in $\alpha$ produces only an intrinsic variation in $M_{bh}$ and that $M_{bh}$ and $S_{bh}$ do not change due to thermodynamic accretion or emission. However, it is {\it not} possible for an $M_{bh} >> M_{pl}$ black hole to be in thermodynamic equilibrium with its environment \cite{H1}: because $T_{bh}\propto 1/M_{bh}$, accretion and emission are always non-zero and $M_{bh}$-dependent for $T_{bh}>0$, and $T_{bh} = 0$ can not be achieved by a finite number of steps. $S_{bh}$ is not even approximately constant in the regime most relevant to the tightest constraints on $de/dt$, i.e. $M_{CMB}>>M_{pl}$. Schwinger $e^+e^-$ pair-production, which \cite{F1} invokes qualitatively, is already explicitly included in \cite{JHM1} and gives rise to the tightest constraint on  $de/dt$. 

Although the argument of \cite{F1} fails, one could conjecture in an extended model intrinsic variation in $M_{bh}$ in addition to accretion and emission (and the second-order accretion and emission rate variations due to $de/dt$). As remarked in \cite{JHM1}, extension of the methodology of \cite{JHM1} to other inter- or independently varying fundamental constants or physics beyond the Standard Model is straightforward. The constraint that the total entropy can not decrease, i.e.~$\Delta S_{total} = \Delta S_{bh} + \Delta S_{env} \geq 0 $, over any $\Delta t$ (which was incorrectly stated in \cite{Davies}) limits the time variation of any parameter, with the terms contributing to the limit and its degeneracy or non-degeneracy potentially differing between models. For intrinsic variation in $M_{bh}$, the calculation of \cite{JHM1} would be supplemented by a ${{dS_{bh}}\over{d \mu}} {{\partial \mu}\over{\partial \alpha}}{{d\alpha}\over{dt}}$ term or any new scalar fields etc.~incorporated in the black hole area $A_{bh}$ definition. While \cite{F1} nonessentially assumes quantized $A_{bh}$ and $S_{bh}$, which depends on unknown Planck-scale behaviour, Eq (3) of \cite{F1} with or without constant $S_{bh}$ implies that, if $A_{bh}$ and $S_{bh}$ are quantized and $\alpha$ varies continuously with $t$, then $\mu$ must be quantized for $Z=0$ but unquantized for $Z \neq 0$. Inversely, if $\mu$ is quantized, $\mu$ can not vary with continuous $\alpha (t)$. Thus it may be more natural to investigate $M_{pl}$ and $M_{bh}$ varying in unison due to a time-varying gravitational constant $G$ as was done in \cite{JHM2}. In QED where the renormalized $e$ depends on the energy scale, the apparent continuous cosmic microwave background cooling implies that either $A_{bh}$ varies continuously when $Z \neq 0$ or the area quanta are extremely tiny.

\end{document}